\documentstyle[osa,manuscript]{revtex}

\begin{document}
\title{Conflict between the Gravitational Field Energy and the Experiments}
\author{Rafael A. Vera}
\address{Departamento de F\'{i}sica. Fac. de Ciencias F\'{i}sicas y Matem\'{a}ticas.\\
Universidad de Concepci\'{o}n. Chile}
\date{}
\maketitle

\begin{abstract}
From the equivalence principle and true gravitational (G) time dilation
experiments it is concluded that ``matter is not invariable after a change
of relative position with respect to other bodies''. As a general principle
(GP), such variations cannot be locally detected because the basic
parameters of all of the 'well-defined parts' of the instruments change,
lineally, in the same proportion with respect to their original values''.
Only observers that don't change of position can detect them. Thus, to
relate quantities measured by observers in different G potentials they must
be previously transformed after Lorenz and G transformations derived from
experiments. They are account for all of the ``G tests''. However ``they are
not consistent with the presumed energy exchange between the field and the
bodies''. The lack of energy of the G field is justified from the GP,
according to which particles models made up of photons in stationary state
obey same inertial and G laws as particle. Such model has been previously
tested with relativistic quantum-mechanics and all of the G tests.
\end{abstract}


%
%
%

%

\section{INTRODUCTION}

The main purpose of the present work is to find, after using ``strictly
homogeneous relationships'', whether or not the current hypotheses normally
used in gravitation are simultaneously consistent with all of the
experimental facts. Obviously, this must be done independently on any
conventional or non-conventional hypothesis.

In current physics, from long time ago, it has been normally assumed that 
{\it the gravitational (G) field exchanges energy with the bodies}\cite
{1Einstein}. Such hypothesis is a consequence of another more basic one on 
{\it the absolute invariability of matter} after a change of G potential
with respect to the observer, which corresponds to one alternative of
interpretation of the Einsteinium's Equivalence Principle (EP).

However such principle has two rather opposed alternatives of interpretation

\begin{description}
\item  I) The bodies are really invariable, which is the conventional
hypothesis, and

\item  II) All of the bodies of a local system change in the same proportion
after identical changes of G potential. Thus an observer moving altogether
with his instrument cannot detect such changes because his standard has
changed in just the same proportion as any other part of the instrument.
Only in this way every ratio in his local system can remain constant.
\end{description}

Notice that in the case II), ``any observer that has not changed of
potential can observe the real changes that have occurred to the bodies
after a change of potential''.

On the other hand, {\it the experiments on G time dilation (GTD), as shown
below,} prove, definitively, that the standard clocks located in different G
potentials run with different frequencies with respect to the clock of an
observer in a well-defined potential.

Then the positive results of the GTD experiments can only be consistent with
the alternative II). 

Since the observers located at rest in different G potentials have standard
clocks running with different frequencies with respect to each other, then
their unit systems are different to each other. Then the current
relationships between the quantities measured by observers in different G
potentials are inhomogeneous and without a well-defined physical meaning. To
relate them, homogeneously, they must be previously corrected for the
differences of frequencies of their clocks. Then the corrected quantities
must be position dependant quantities.

Thus, the new formalisms that must be used, to describe the real physical
changes of a body, after a change of its position in a G field, is a plain
generalization of the one used in special relativity, after using position
and velocity dependant quantities.

As a matter of fact, the present work is based almost exclusively on the G
Time Dilation (GTD) experiments and the Equivalence Principle (EP). The full
consistency with all the other experiments can be verified later on

\section{ THE EXPERIMENTAL FACTS}

For the purposes of the present work, the experiments in G fields have been
divided into two main categories whose general results have some fundamental
differences with respect to each other.

\begin{description}
\item[I. ]  {\it Local experiments} in which the differences of G potential
between the object and the observer are small enough so that special
relativity can be applied locally.

\item[II. ]  {\it Nonlocal }(NL){\it \ experiments} in which the objects and
the observers are located in different G potentials.
\end{description}

\subsubsection{The global result of local experiments}

The EP is already a general result from the most exact local experiments.
According to it, {\it the special relativistic forms of the local physical
laws do not change after any well-defined change of velocity and G potential
of the observer and his measuring system }\cite{3Misner}. Thus, ``{\it the
``ratios'' between the local parameters of the atoms and particles are
constant values that do not depend on the velocity and on the G potential of
the local system}''. For this reason, and for strictly ``local purposes'',
every local observer arbitrarily ``{\it assigns}'' the same numerical value
to the frequency of his local (standard) clock. According to the EP, such
assigned frequency also fixes the rest values of the other the parameters of
the local system. Then, such constant number can only be legally used for
strictly ''local'' relationships in which the differences of G potential
between the object and the observer, can be neglected\footnote{%
The local case is similar to that of a small country. The number ``1''
assigned for the local moneys cannot be legally used for buying in other
countries.}.

\subsubsection{The global results of the non-local experiments}

The results of the most exact experiments used for testing gravitational
theories, currently known as ``gravitational tests'', show most clearly,
that some real ``physical changes'' do occur to the bodies and to the space
after a change of position in a G field. For example, the deviation of light
in a G field gradient proves, definitively, that the G field has a variable
refraction index with respect to any well-defined observer.

The physical changes occurring to the bodies are most evident in the
experiments on G Time Dilation (GTD) made up with clocks. They have
important differences with the experiments on G Redshift (GRS) of photons
because the observed time intervals between two pulses of light do not
depend either on the frequency of photons or on their times of flight
between the clock and the observer. A time interval is a difference of times
in which such time of flight is cancelled out. The frequency of such photons
is not measured at all. Then the GTD experiment is entirely independent the
mechanisms of propagation of light signals between the clock and the
observer.

This experimental facts are most clear when the readings of atomic clocks
travelling in airplanes or satellites have are compared. For example the
Cesium clock of the Global Position System (GPS) NTS-2, during a period of
20 days of June of 1977, accumulated a difference of time of the order of
760,000 nanosecond. The average frequency of the NL clock measured during
that interval was +442.5 parts in 10$^{12}$ faster than clocks on the
ground, which is in good agreement with the GTD predicted by general
relativity.

Then the GTD experiments prove, definitively, that {\it the clocks located
in different G potentials do run with different rates with respect to each
other, i.e., that the objects are not absolutely invariable after a change
of G potential.}

On the other hand, from the change of sign observed after exchanging the
position of the object with the observer it is concluded that: the
differences of frequencies are {\it absolute} ones.

Since the eigen-frequency of any clock depends on the general physical
properties of its atoms, then a general conclusion derived from GTD
experiments is that{\it : }``{\it the standard atoms of observers located in
different G potentials are physically different with respect to each other,
respectively''}.

Then, from the positive results of the GTD experiments it is inferred that, 
{\it the current comparisons of frequencies measured with clocks located in
different potentials are inhomogeneous and without well-defined physical
meaning. They are referred to reference clocks that run with different rates
with respect to each other}. Such illegal kind of relationship is bounded to
be a current source of fundamental errors similar to the classical ones that
existed before Einstein. Since this is a current practice, then it is
reasonable to find some fundamental errors in current literature.

Consequently, to fairly relate the quantities measured by observers at rest
in different G potentials, after strictly homogeneous relationships, all of
the quantities must be previously transformed to some strictly invariable
unit system based on a clock or reference standard located at rest in some
well-defined potential. This is because only such kind observer can be
strictly invariable. For the present purposes, such transformations must be
found from the experimental facts, independently on any of the conventional
hypotheses.

\subsubsection{Global Results from Local and Nonlocal Experiments}

According to the EP, {\it the constants relating the frequencies, the masses
and the lengths, of any well-defined part of a local system, are values that
do not depend on the velocity and G potential of such system}. Thus if any
of these parameters change, after a change of potential with respect to a
fixed observer, then the other parameters must also change in just the same
proportion. Only in this way every local ratio within the instrument can
remain unchanged.

Then the GTD experiments can be consistent with the EP only if:

{\it The basic parameters of any well-defined part of any measuring system
change linearly, in just the same way and in the same proportion, after any
common change of G potential}\footnote{%
Only in this way every ratio within the measuring system can remain
unchanged after any circumstance. Notice that if this were not so, the
differences could be detected from local measurements thus violating the EP.}%
.

Then according to the EP and to the positive results of the GTD experiments,
an observer travelling altogether with his local instruments cannot detect
any change of the local ratios, after a change of G potential, because all
its parts change in the same way and in the same proportion, after identical
changes of G potential''. The reason is obvious: ``All of the well-defined
parts of the instrument obey the same inertial and gravitational laws and,
therefore, they must change in just the same way and in the same proportion
after identical circumstances''. This may be called {\it the Nonlocal form
of the EP} (NL EP).

Since a similar fact is observed in special relativity, after velocity
changes, then the common point for the two cases is that:

{\it All of the well-defined parts of the measuring system, without any
exception, obey the same general inertial and gravitational laws\footnote{%
{\it If the inertial or gravitational laws of some particle were different
from another particle, such differences could be detected after local
measurements made up after changes of velocity and G potentials of the
measuring system. Such positive results would violate the EP.}}. }This is 
{\it a more general principle (GP)} because it is a general condition that
the bodies must meet to account for all of them: the EP, special relativity
and the G experiments.

It is important to observe that the same principle must also hold for any
photon in stationary state between two well-defined parts of the same
system. This is because its frequency and its wavelength have well defined
values fixed by the local speed of light. Its mass-energy is fixed by its
frequency, according to $m=E=h\nu $. Thus, an idealized clock can be
emulated by a photon in stationary state between two well-defined parts of
the same system.

\section{The new formalism for an homogeneous description of the
gravitational phenomena}

From the changes of frequencies of the clocks revealed by the GTD
experiments it may be concluded that:

\begin{itemize}
\item  {\it The constant numerical values assigned to the rates of the local
clocks are well defined (legal) only for strictly local relationships in
which the GTD phenomenon can be neglected.}

\item  {\it In principle the relative changes occurring to the objects,
after changes of G potential can only be measured (or described) by
observers that have not changed of G potential}.

\item  {\it For a complete description of all of the phenomena occurring in
G fields it is essential that the reference standard has not had any of the
changes that have occurred to the objects. This one fixes a strictly flat
theoretical reference frame.}

\item  {\it To relate quantities measured by observers in different G
potentials, they must be previously transformed to some common reference
standard (Lorenz frame) in some well-defined state of velocity and G
potential. This is equivalent to use a strictly invariable (flat) reference
framework that has not changed in the same way as the objects.}
\end{itemize}

Here, the basic relationships for transforming\ the data of observers at
rest in different positions of a G field, to some common observer at rest in
some well-defined position, are called ``{\it G transformations}''.

Below, {\it the transformation factors are derived from the Equivalence
Principle and the data of GTD experiments after using strictly homogeneous
relationships.} Those factors correct each quantity for the real physical
changes that have occurred to the nonlocal bodies and clocks after changes
of G potential with respect to the observer's ones.

In more general cases, when the NL bodies are moving with respect to the
observers the ``{\it Lorenz Transformations}'' can be as important as the
Gravitational Transformations. In such cases {\it the two kinds of
transformations factors must be applied}. Using can do this

a), {\it Lorenz Transformations factors} for the description of the relative
changes that the bodies have had after changes of velocity, and

b), {\it Gravitational Transformations}, for the description of the absolute
changes occurring to the bodies after changes of G potential..

In this way the relative and absolute differences between the NL bodies and
the local ones, due to differences of G potential and velocity, have been
taken into account.

The last transformations extend the application range of special relativity
for nonlocal cases in G fields. Thus a reasonable name for this formalism is 
{\it nonlocal relativity.}

For the present purposes it is obvious that it not necessary to use the
geometry of space-time. However the present formalism gives a direct and
complete (non-distorted) description of the net changes occurring both in
matter and in the space, after changes of position and velocity with respect
to some fixed (invariable) reference frame.

\subsubsection{Conventions}

To prevent ambiguities, it is necessary to make clear the differences
between the {\it local} and the {\it nonlocal} conditions of the object with
respect to the observer, depending on whether or not the objects are in the
same G potential of the observer. Thus the terms ``{\it NL objects}'' and ``%
{\it NL observers}'' are used here for the nonlocal conditions. For the same
reason, the transformed quantities, after correction for differences of
velocity and G potential, have been named ``{\it NL quantities}''. They
correspond to a plain generalization of the ``{\it relativistic quantities'',%
} used in special relativity.

Here, for simplicity, a static central field has been used. In this way the
G potentials are plain (point) functions of just the radius. The observer's
standard is located at rest in some fixed radius of some central field,
which is assumed to be strictly static. However, due to its high importance,
most of the times {\it the fixed position of the observer has been
explicitly stated in each quantity by means of a ``subscript''.}

For example, it is assumed that some observer $A$ and his standard clock are
located in some NL radius $a_{a}$ of a central body of nearly infinite NL
mass $M_{a}$ compared with the test mass$.$ Another standard clock is
located in some NL radius $r_{a}$. Notice that the common subscripts ($_{a}$%
) puts into relief that these radii are expressed in terms of the common
unit system fixed by the standard of the observer $A$.

The GTD experiments put on relief that the frequency of a clock ``at rest''
with respect to the observer is a well-defined (point) function of the NL
radius of the NL clock ($r_{a}$) and of the observer's clock $A$ ($a_{a}$).

On the other hand, according to special relativity, if the NL clock were
moving with respect to the observer, its NL frequency would also be a
function of its velocity ($V_{a}$) with respect to the observer $A$. Thus
the general symbol used here for the NL frequency of such clock, with
respect to the observer $A$, is $\nu _{a}(V,r)$. The same symbol is used for
the average NL frequency of any well-defined standing wave of the NL system.
Notice that, for simplicity, the subscripts of the variables in parenthesis
are omitted. However, by convention, it is assumed that the tacit subscripts
are the same for all of the variables of the same quantity.

For an object ``at rest'' with respect to the observer, the value of the
velocity ($V=0$) is ``not'' omitted. Thus the symbols $\nu _{a}(0,a)$ and $%
\nu _{a}(0,r)$ are used for the frequencies of a local clock and a NL one,
respectively. They are at rest with respect to the observer {\it A}.

According to current conventions, the numerical values of assigned to the
rates of the ``local'' clocks are universal values that do not change when
the observer changes of G potential. For this reason in the current
literature most of the times the positions and velocities of the observer
are not explicitly stated. However, according to the GTD experiments, such
rate has really changed after a change of G potential. Then, as shown in the
discussion (below), such practice applied to relations between quantities
measured in different G potential has been source of fundamental errors, as
shown below. For this reason, the observer and the object positions must
also be explicitly stated.

Due to the different frequencies of the clocks located in different
positions in the field, the NL speed of light is also position dependent.
This is consistent with the refraction phenomena observed in G fields. Thus
the in the case of {\it a free photon in a NL position }$r$, its basic NL
parameters with respect to the observer have the symbols $\nu _{a}(r)$, $%
\lambda _{a}(r)$. The NL speed of light, with respect to the observer $A$, $%
c_{a}(r),$ can depend only on the NL positions of the photon and of the
observer. This is not in conflict with the constant value of the ``local''
speed of light because the differences between the NL values and local
values tend to zero when the differences of G potential between the object
and the observer tend to zero. Thus when $r_{a}\rightarrow a_{a}$, $%
c_{a}(r)\rightarrow c$.

For a photon in stationary state, in a NL system at rest with respect to the
observer, its NL speed of light, is obviously equal to:

\begin{equation}
c_{a}(r)=\nu _{a}(0,r)\lambda _{a}(0,r)  \label{1}
\end{equation}

In which $\nu _{a}(0,r)$ and $\lambda _{a}(0,r)$ are the average values of
the NL frequency and NL wavelength of the radiation in stationary state with
respect to the observer.

\section{GRAVITATIONAL TRANSFORMATIONS FROM EXPERIMENTS}

Here the G transformations are derived from just experimental facts. Using
the EP, which is already a global result of the local experiments, and the
results of GTD experiments, can do this. In this way the simultaneous fit of
the G transformations with the rest of the gravity tests can be used as
experimental ``tests'' for such transformations.

Assume a GTD experiment in which a standard clock is raised from the NL
radius $r_{a}$ of the observer $A$ up to a NL radius $r_{a}^{^{\prime
}}=r_{a}+dr_{a}$. If $\nu _{a}(0,r)$ is the initial frequency with respect
to the observer $A$, the experiments show that the proportional change of
the frequency of the NL clock, compared with the original frequency has the
form:

\begin{equation}
\frac{\nu _{a}(0,r+dr)-\nu _{a}(0,r)}{\nu _{a}(0,r)}=d\phi (r)\cong \frac{gdr%
}{c^{2}}\cong \frac{GM_{a}}{r_{a}}\frac{dr_{a}}{r_{a}}=\frac{dE_{a}}{%
m_{a}(0,r)}  \label{2}
\end{equation}

In which the last members are the experimental values in which, for
simplicity $M[joule]=M^{newt}c^{2}$, and $G=G^{newt}c^{-4}$. In this way the
mass and the energy can be expressed in terms of a common mass-energy unit
(joule).

On the other hand, a{\it ccording to the EP, }the local values of the
frequencies, mass-energies, wavelengths and lengths of any well-defined part
of a local system ARE related to each other after some relation like:

\begin{equation}
k^{\nu }\nu _{a}(0,r)=k^{m}m_{a}(0,r)=k^{\lambda }\lambda
_{a}(0,r)=k^{L}L_{a}(0,r)\text{ }  \label{3}
\end{equation}

In which the constants $k^{\nu }$, $k^{m}$, $k^{\lambda }$, and $k^{L}$ do
not change after changes of velocity and G potentials. Thus {\it if any of
these parameters change, after a change of G potential with respect to the
fixed observer, the other ones must also change in just the same
proportions, compared with their original values.}

Then, from equations \ref{2} and \ref{3}, the common NL changes occurring in
any part of a system, after a change of G potential, can be represented by
the single expression :

\begin{equation}
\frac{d\nu _{a}(0,r)}{\nu _{a}(0,r)}=\frac{dm_{a}(0,r)}{m_{a}(0,r)}=\frac{%
d\lambda _{a}(0,r)}{\lambda _{a}(0,r)}=\frac{dL_{a}(0,r)}{L_{a}(0,r)}=d\phi
_{a}(r)=\frac{dE_{a}}{m_{a}(0,r)}  \label{4}
\end{equation}

This means that such NL parameters of the bodies are no longer the constant
(universal) values. All of them are well-defined point functions of the NL
positions of the object and of the observer with respect to the source of G
field

In particular, from the NL\ EP, this relation must hold for the frequency
and the wavelength of any standing wave in the NL system at rest with
respect to the observer. In this case , from \ref{1} and \ref{4} 
\begin{equation}
\frac{d\nu _{a}(0,r)}{\nu _{a}(0,r)}=\frac{d\lambda _{a}(0,r)}{\lambda
_{a}(0,r)}=\frac{1}{2}\frac{dc_{a}(r)}{c_{a}(r)}  \label{5}
\end{equation}

In a more general case, when a local instrument is changed from the position
of the observer $A$ up to a NL radius $r_{a}$, the common G transformation
factor can be derived integration of \ref{4} and \ref{5}:

\begin{equation}
f_{a}(r)=\frac{\nu _{a}(0,r)}{\nu _{a}(0,a)}=\frac{m_{a}(0,r)}{m_{a}(0,a)}=%
\frac{\lambda _{a}(0,r)}{\lambda _{a}(0,a)}=\sqrt{\frac{c_{a}(r)}{c_{a}(a)}}%
=e^{\Delta \phi _{a}(r)}\cong 1+\Delta \phi _{a}(r)  \label{6}
\end{equation}

{\em The first four members} of equation \ref{6} define {\it the G
transformation factor} between the NL and the local parameters of each
well-defined part of the same system at rest in the two places. From the EP,
the same ratio must hold for any clock, atom, particle, or any well-defined
radiation in stationary state that may exist at rest in the local and NL
systems. Thus the phenomenon of G redshift (GRS) of a photon coming from a
NL atom is a consequence of the GTD that existed in the atom before emitted
it .

{\em The fifth member} gives the explicit values of the transformation
factors in terms of the NL speed of light. This member puts on relief that 
{\it the G field is a gradient of the NL refraction index of the space}.

{\em The last member of equation }\ref{6} is the experimental value found
for the 2nd member, according to single GTD experiment.

{\em The sixth member} is the NL transformation factor obtained by
integration of the values obtained from GTD experiments.

From equation \ref{6} it is verified that the parameters of a NL object at
rest with respect to the observer\ are point functions that depend on the NL
positions of the object and of the reference standard. Something similar
occurs for the NL speed of light.

Notice that, in general, {\it in regions of lower NL speed of light there is
a general contraction of the rest-values of all of them: the frequencies,
the mass-energies, the wavelengths and the lengths}. This holds for every
well-defined particle or radiation in stationary state.

\subsection{Gravitational tests}

The equation \ref{6} is obviously consistent with the EP because in any
local limit $\phi _{a}(a)=1$. This value is independent on velocity and G
potentials of the local system. This means a full agreement with the most
exact experiments.

In previous works the equation \ref{6} has been derived in an entirely
different way, from the theoretical properties of a particle model made up
of radiation in stationary state. In them it has been proved that {\it the
different members of this equation are consistent with all of the
experiments used for testing gravitational theories }\cite
{4Vera81a,5Vera81b,6Vera97}.

For example, the same transformation factor has been found from the GRS
experiments \cite{7Pound}. In them, the frequency of light coming from a NL
luminous source, located in a different G potential, has been measured. In
them, the NL form of the EP is clearly verified because ``{\it the
frequencies of the clocks and of all of the spectrum lines of the atoms turn
out to be redshifted in just the same proportion''}.

It is simple to verify that the gradient of the NL speed of light fixed by
the 5th member accounts for the results of the experiments on\ {\it the time
delay of radar waves traveling close to the Sun }\cite{8ShafRad}. {\it This
member also accounts for the deviation of light travelling close to the Sun} 
\cite{9BertotiRef}{\it .}

In such literature it is proved that the theoretical orbit of a particle
model, derived from equation \ref{6}, is also consistent with\ the observed
perihelion shift of Mercury \cite{10ShafPeri}. In this case the perihelion
shift depends on a second order approximation of the 6th member this
equation.

Then it may be said that ``{\it equation }\ref{6}{\it \ is simultaneously
consistent with all of the local and NL experiments made up in G fields}''.
Since this one does not depend on any particular G theory, then it may be
used to test the current hypotheses normally used in G theories.

\section{CONFLICTS OF SOME CURRENT HYPOTHESES WITH THE EXPERIMENTS}

Some of the most basic questions are:

\subsection{Can a photon exchange real energy with G fields?}

Assume that several experiments on G redshift are made up by different
observers located consecutively along the trajectory of a light beam going
away from a central body. Such observers would find ``decreasing frequencies
values''. This fact currently makes believe in that light is redshifted
during its way out and, therefore, that there is a real exchange of energy
between the photons and the field. Thus it is often assumed that the photons
Is doing a work, which is the {\it tired light hypothesis}.

However, such reasonings are inhomogeneous because, from GTD experiments,
such frequencies are referred to clocks that are running with different
frequencies. Then such differences of frequencies have no well-defined
physical meanings.

Then the single way to compare such frequencies is after transforming the
frequencies, previously, to some well defined clock located in some
well-defined G potential. This can be done, by using the equation \ref{6}
obtained above from GTD experiments or any other gravitational test.
According to it the photons emitted by the NL oscillators at $r_{a}$ should
have an initial NL frequency, with respect to the observer A, given by:

\begin{equation}
\nu _{a}(r)=\nu _{a}(0,a)\left[ 1+\Delta \phi _{a}(r)\right]  \label{7}
\end{equation}
If the field had given up some energy $\Delta E$ to the photon, the final
frequency of the photon, after travelling up to the observer's potential,
would be somewhat higher, like :

\begin{equation}
\nu _{a}(a)=\nu _{a}(0,a)\left[ 1+\Delta \phi _{a}(r)\right] +\Delta E/h
\label{8}
\end{equation}

On the other hand, according to the results of the experiments on G
redshift, the value of the frequency at the observer potential is given by

\begin{equation}
\nu _{a}(a)=\nu _{a}(0,a)\left[ 1+\Delta \phi _{a}(r)\right]  \label{9}
\end{equation}

Then, after comparing \ref{7}, \ref{8} and \ref{9}, it is concluded that

\begin{equation}
\Delta E=0\quad ;\quad \nu _{a}(r)=\nu _{a}(a)  \label{10}
\end{equation}

This means that, within the experimental errors,

\begin{itemize}
\item  The G field does not exchange energy with photons. This rules out the
tired light hypothesis.

\item  The NL frequency of photons, with respect to a clock in a fixed G
potential, remains constant during its trip through the ordinary gradients
of G fields (NL frequency conservation of free photons).

\item  The presumed energy exchange between the photons and the G field is
not simultaneously consistent with the GTD and GRS experiments.
\end{itemize}

\subsubsection{Verification from ordinary properties of light}

\begin{quote}
According to ordinary experiments, a gradient of the refraction index does
not change the frequency or color of the radiation, i.e., photons do not
exchange energy with the dielectric. According to equation \ref{6}, light is
deviated in a G field according to normal refraction laws. Consequently, ``%
{\it photons do not exchange energy with static G fields}''.
\end{quote}

\subsubsection{Verification from wave properties of light}

\begin{quote}
From current experiments made up with of single photons it is inferred that
even {\it single photons have wave properties}. Then {\it each photon }must
be the result of constructive interference of a large number of {\it %
``wavelets''}. Thence the ``{\it wave}-continuity'' of a light beam must be
a direct consequence of the ``wavelet continuity''. According to such
``continuity'', a change of the NL refraction index of the space would not
change the net number of waves of a wavelet-train. It would produce just a
change of its NL wavelength {\it without a net change of its NL frequency}.
It is important to observe that the different observers along the trajectory
of a wave train should observe {\it the same number of waves}. However they
would observe different rates because their clocks run with different
frequencies.

This may be even more clear if pulses of light and {\it bullets} are sent
after well defined number of waves of electromagnetic radiation. Since the
number of pulses or bullets cannot change, regardless of the current work
(tired light) hypothesis, then single alternative for explaining the
different number of bullets per second detected by observers located in
different potentials is that their clocks run with different rates.
\end{quote}

Then it is concluded, either from experiments or from general properties of
radiation, that:

\begin{itemize}
\item  {\it Photons do not exchange energy with strictly static G fields}%
\footnote{%
Curiously, the NL frequency conservation law can be demonstrated with the
aid of the results of GRS experiments, which are often interpreted in the
opposite sense.} (The no exchange law for G fields.)

\item  \ {\it The differences of frequencies observed in the GRS experiments
are due to differences of the eigen-frequencies of the atoms and clocks
located in different G potentials}. {\it The G redshift phenomenon has
occurred in the bodies, ``before'' the emission of light}. It does not occur
during the light trip.
\end{itemize}

\subsection{Does a G field exchange energy with the bodies?}

Let us find, after using strictly homogeneous relationships, whether or not
such hypothesis can be ``simultaneously'' consistent with the Equivalence
Principle and with the gravitational experiments.

A direct answer comes from the 2nd and the last member of Eq. \ref{4}, from
which $dE_{a}=dm_{a}(0,r)$. Thus the energy used up to raise the body is not
stored in the field but in the body, as an extra mass-energy. The same
result can be derived from other ways, as follows:

\subsubsection{The no exchange law for bodies derived from free fall
experiments}

Assume that an observer $A$ is located at rest in a fixed radius $a_{a}$ of
a central field. He observes, locally, the free fall of a test body which
was initially at rest at $B$ in some NL radius $r_{a}=a_{a}+h_{a}$ .

For a real mass-energy balance during the free fall, the initial and final
values of the NL rest masses, both referred the unit system of the observer $%
A$, must be known with high exactitude.

Strictly, the observer at $A$ does not know the exact initial (NL) rest mass
of the body at $B$, compared with its final rest mass at $A$. However he can
find it either from GTD experiments done before the free fall, i.e., after
using the transformations given in equation \ref{6}. According to it, the
initial (NL) mass of the test body at $B$, with respect to the observer $A$,
is:

\begin{equation}
m_{a}(0,r)=m_{a}(0,a)\left[ 1+\frac{GM_{a}}{r_{a}}\frac{h_{a}}{r_{a}}\right]
\label{11}
\end{equation}

On the other hand, according to special relativity and the current results
of free fall experiments, the final (local) mass of the body at $A$, just
``before the stop'', is.

\begin{equation}
m_{a}(V,a)=\gamma m_{a}(0,a)=m_{a}(0,a)\left[ 1+\frac{GM_{a}}{r_{a}}\frac{%
h_{a}}{r_{a}}\right]  \label{12}
\end{equation}

Notice that this is a strictly local (legal) relationship. Then, from \ref
{11} and \ref{12}, the single strictly homogeneous relationship between the
initial and final mass-energies during the free fall is:

\begin{equation}
m_{a}(0,r)=m_{a}(V,a)  \label{13}
\end{equation}

On the other hand, the net energy released during the stop is:

\begin{eqnarray}
\Delta E_{a} &=&m_{a}(V,a)-m_{a}(0,a)=m_{a}(0,r)-m_{a}(0,a)  \label{14} \\
&\cong &\frac{GM_{a}m_{a}(0,a)}{r_{a}^{2}}h_{a}  \nonumber
\end{eqnarray}

From equations \ref{13} and \ref{14} it may be concluded that

\begin{itemize}
\item  {\it During a free fall, the mass-energy of a body, with respect to
some well-defined observer, remains constant (NL mass-energy conservation
during a free fall.).\ }

\item  {\it The net energy released from G work comes not from the G field.
It is just a small fraction of the rest mass-energy of the test body, which
is liberated during the G work.}

\item  {\it There is not a true exchange of energy between the field and the
test body (The no energy-exchange law for G fields.).}
\end{itemize}

\subsubsection{The no energy-exchange law derived from gedanken-experiments}

{\it Experiment 1. Matter and anti-matter annihilation occurring during a
free fall.}

Assume an observer at rest far away from a neutron star and an electron pair
falling vertically into a neutron star. Statistically the annihilation may
occur in any radius. Assume, for simplicity, that the pair decays into two
gamma photons travelling in symmetrical trajectories with respect to the
annihilation radius, so that the two photons have the same energy with
respect to the observer.

According to global mass-energy conservation made up by the observer at
infinity, {\it the net energy going far away from the system cannot depend
on the specific radius in which annihilation occurs}. Then the ``NL
annihilation'', occurring during the fall must produce the same net energy
with respect to the observer, regardless on the radius in which it has
occurred.

\begin{quote}
\begin{equation}
m_{\infty }(V,r)=2h\nu _{\infty }(r)=2h\nu _{\infty }(\infty )=m_{\infty
}(0,\infty )  \label{15}
\end{equation}
\end{quote}

The NL annihilation, stated in 1st and 2nd member of equation \ref{15}, must
be equal to the local one occurring far away from the system, stated in the
3rd and 4th member. The conservation of the NL frequency of the photons
travelling between $r_{\infty }$ and $\infty $ is obvious from the 2nd and
3rd members, and the one for the NL mass of the particles is obvious from
the first and last members.

Then it may be concluded that {\it during the free fall of the particle-pair
its NL mass-energy remains constant. The same holds for the NL energy of the
photons.}

{\it Experiment 2. Radioactive decay occurring during a free-fall.}

A similar gedanken-experiment can be done by assuming the free fall of a
radioactive atom that may decay producing a gamma photon in any arbitrary
radius of a central field.

According to the EP, {\it the energy of a photon is a constant fraction of
the mass of an atom}. This fraction is independent on the radial position in
which the atom decays. Thus, for an observer at $\infty $, this fraction
should be the same both for the local position ($\infty $) and for any other
NL position($r_{\infty }$) during the fall. This means that:

\begin{equation}
\frac{\Delta E}{m}=\frac{h\nu _{\infty }(\infty )}{m_{\infty }(0.\infty )}=%
\frac{h\nu _{\infty }(r)}{m_{\infty }(V.r)}=Constant  \label{16}
\end{equation}

According to NL frequency conservation, given in \ref{10}, the numerators of
the equation \ref{16} must be the same. Consequently, the denominators of
such equation must also be the same values. This means that ``{\it during
the free fall of the atom, its NL (relativistic) mass-energy with respect to
some well-defined the observer, is conserved''}.

\section{THEORETICAL DISCUSSION OF THE EXPERIMENTAL RESULTS}

To understand the nature of the G phenomena we must understand first the
nature of matter. Some help can be found, directly from the NL EP.

\subsection{Help from the minimum well-defined ``part'' of a system.}

According to the NL form of the EP, found above from all of the experimental
facts, {\it all of the well-defined parts of a local system must obey the
same inertial and gravitational laws}. Then the same should hold for ``the
minimum well-defined part of a local system''. Thus the general properties
of uncharged particles can be understood in terms of the minimum
well-defined part of a system, which is ``{\it a single quantum of radiation
in stationary state''.} This quantum may be confined between any other
well-defined parts of a systems, like a wave cavity. The last one fixes
well-defined values of the wavelength, frequency and mass-energy of the
quantum in stationary state\footnote{%
This kind of ``particle model'' can also be found after consecutive
simplification of the Michelson-Morely experiments and of Kennedy's ones\cite
{11Kennedy} made up with arms of different lengths. From the negative
results of both kinds of experiments it may be inferred that the number of
wavelengths on each arm must remain unchanged after changes of velocity and
G potential. Thus the length of the rod and the wavelengths of the radiation
must change in the same proportion after identical circumstances.}.

Then it may be concluded that, according to the EP, {\it the general
properties of uncharged particles and their fields can be derived from the
general properties of particle models made up of a photon in stationary
states}.

Notice that, the net number of parameters of the particle model at rest with
respect to the observer is minimum\footnote{%
For the present purposes, the values of spins or other internal details of
the model can be ignored.}. For a model in a potential different from the
observer's one, they are its NL frequency with respect to the observer $A$,
called $\nu _{a}(0,r)$, and its NL wavelength, $\lambda _{a}(0,r)$. The
product of such variables fixes the value of the NL speed of light,
according to equation \ref{1}, or vice versa. Its NL mass-energy with
respect to the observer $A$ is, by definition, the energy confined in it:

\begin{equation}
m_{a}(0,r)=h\nu _{a}(0,r)  \label{17}
\end{equation}
Thus the mathematics involved in this approach turns out to be extremely
simple\footnote{%
In previous works \cite
{4Vera81a,5Vera81b,6Vera97,12Vera83,13Vera86,14Vera96a,15Vera99a}, it has
been found that the particle model provides a direct way to account for the
transformations given by equation \ref{6}. It has also been used to find
self-consistent explanations for basic relations in special relativity,
quantum mechanics and gravitation. Thus the consistency of the properties of
the particle model and those of any other uncharged particle seem to be
fairly well tested.}.

Consequently, {\it the particle model can be used to find more fundamental
reasons for the above results}.

For example, it may be easily verified that {\it such particle model cannot
change of velocity unless that a gradient of the NL refraction index exists
in the space}, i.e.{\it , without the G field gradient shown in Eq. (4).}

\subsection{Why G fields do not exchange energy with the particles?}

The physical reasons for the above results can be understood from several
different viewpoints.

\begin{description}
\item[a) ]  {\bf \ Explanation according to the nature of the particle model}
\end{description}

According to Eq. (6) the main phenomenon producing the G acceleration is
just ``{\it refraction}''. For a transversal model, for example, the
increase of momentum comes from the net deviation of the trajectories of
light occurring during a round trip of the wave propagation inside it. It is
well a well-proved fact that the phenomenon of refraction ``{\it does not
change the frequency of the photons}'', i.e., it does not exchange energy
with the dielectrics.

Then the particle model does not exchange energy with the field because it
is made up of photons that do not exchange energy with strictly conservative
fields.

A more detailed explanation on the mechanism of acceleration of the particle
model in a static G field is given in the references \cite{5Vera81b,6Vera97}

.

\begin{description}
\item[b) ]  {\bf Explanations according to the nature of the G field.}
\end{description}

{\it The G field of a particle model can only depend on long-range
properties of a single photon in stationary state.}

Such properties can be learned from the results of optical experiments made
up with single photons. From them it is simple to conclude that:\ 

\begin{itemize}
\item  The wave properties of a single photon should come from the wave
properties of more elemental kinds of ``{\it wavelets}''. They must be
interfering constructively in the photon and destructively far away from it.

\item  The wavelets are not destroyed during destructive interference with
other wavelets. Thus, in other places, they interfere with other wavelets
independently on their previous interference's with other wavelets\footnote{%
The wavelets associated to single photons have also been used, in the works
cited above describe other physical properties of uncharged particles.}.

\item  A photon in stationary state between dielectric mirrors must be the
result of interference of wavelets travelling in opposite directions. They
must interfere constructively only within its reflection zones\footnote{%
In optics, the existence of such reflection zone, outside of the dielectric,
is obvious from the experiments on ``frustrated reflections''}. Outside of
such zones, they must interfere destructively.

\item  The wavelets going away from a particle model must propagate
themselves rather indefinitely the space with random phases.

\item  The NL properties of G field can only depend on the gradient of the
NL perturbation rate of the space\ produced by{\it \ ``wavelets with random
phases'' }that are actually crossing it.

\item  ``{\it The G field has no energy'' because the net amplitude of
wavelets with random phases is zero}''. Thus, statistically, the probability
for the existence of energy in such space must be zero.
\end{itemize}

\subsection{\ Where the current errors may come from?}

When Einstein conceived his theory on General Relativity he postulated a
field equation by assuming, as a basic hypothesis, that ``{\it the G field
transfers energy and momentum to the matter in that it exerts forces upon it
and give it give it energy}'' \cite{1Einstein}. To support such hypothesis,
he used arguments based on the properties of electric fields. He ignored the
fact that the geometrical and physical properties of electric fields are
radically different to the G ones.

Notice that in the above statement, due to the lack of more convincing
arguments, Einstein tacitly postulated twice the same hypothesis, in
different words. Because to exert a force, or to give up momentum, it is not
necessarily associated to an energy transference. He tacitly ignored the
alternative of the self-propelled bodies that use up their own internal
energies to accelerate themselves after the momentum given up by some static
external force.

Such hypothesis of Einstein has also been supported by the classical-like
hypothesis in that matter is absolutely invariable after a change of G
potential with respect to the observer's one''. According to such hypothesis
the standard clocks of all the observers in different potentials are
physically identical with respect to each other. For this reason, the
position of the observer is not usually stated in the equations of the
current literature.

However such hypothesis is not consistent with the GTD experiments that
prove, as shown here, that some fundamental changes have occurred to the
clocks after change of G potential. Then such experiments can be consistent
with the EP only if all of the parts of a measuring system have ''changed''
in just the same proportion after identical changes of potential. Only in
this way every ratio within the measuring system can remain constant. Only
in this way any local observer moving altogether with the system cannot
detect such changes. In principle, only the observers that have not changed
of potential can observe the real changes that have occurred to the bodies
after a change of G potential. This accounts for the experiments on GTD and
the other G tests.

The errors in current literature can be easily found when the position of
the observer is stated by a subscript. For example: assume that a body falls
freely between the radii $b_{a}$ and $a_{a}$. Assume that the final mass
relativistic mass at $a_{a}$, before the stop, is called $m(V)_{final}$, and
that the initial mass at $b_{a}$ is called $m(0)_{initial}$. Then a current
mass-energy balance for a ``free fall'' would have the form:

\begin{equation}
m(V)_{final}=m(0)_{initial}+\Delta E  \label{18}
\end{equation}

This form of equation makes believe in that $\Delta E$ is some energy that
the field gives up to the body. This one may look perfect, but the lack of
homogeneity of this equation stands out when the positions of the initial
object and the observer are explicitly stated according to the above
conventions:

\begin{equation}
m_{a}(V,a)_{final}=m_{b}(0,b)_{initial}+\Delta E_{a}  \label{19}
\end{equation}

The different subscripts of the different terms of this equation put on
evidence that this equation is an inhomogeneous mixture of quantities
measured by observers in different G potentials. Then {\it this equation has
not well-defined physical meaning because the reference standards of the
observers at }$b${\it \ and at }$a${\it \ are physically different with
respect to each other}.

It may be argued that in equation \ref{19} the value of $%
m_{b}(0,b)_{initial} $ can be replaced by $m_{a}(0,a)_{after\text{ }stop}$
because they have the same numerical values. Effectively, such replacement
gives:

\begin{equation}
m_{a}(V,a)_{final}=m_{a}(0,a)_{after\text{ }stop}+\Delta E  \label{20}
\end{equation}

But now this relation has a radically different meaning. It is just a
mass-energy balance occurring ``during the local-stop period'', which{\it \
is independent on where the free fall energy comes from}. Then it is not
possible to find, from just Eq. \ref{20}, where the G energy comes from.

\section{CONCLUSIONS}

From the EP and the positive results of the GTD experiments it is concluded
that:

\begin{itemize}
\item  Some absolute changes do occur to every well-defined part of any
local system after a common change of G potential. Such changes cannot be
detected from local measurements because the changes of frequencies, masses
and lengths of each part of any local system, occur in just the same
proportion. {\it Such changes can only be detected, and described, by
observers that have not changed in the same way as the objects}.

\item  The reference standards of observers located in different G
potentials are physically different with respect to each other, regardless
of the identical numbers normally assigned to them. Thus must of the current
relations between quantities measured in different potentials are
inhomogeneous and meaningless.

\item  The single way for a complete description of all of the changes that
have occurred to the NL objects, after a change of G potential, is after
using some well-defined (flat) reference frame that has not changed in the
same way as the objects.

\item  To relate quantities measured in different G potentials they must be
previously transformed to some common Lorenz frame in some fixed and
well-defined G potential.

\item  In principle, and in fact, the G transformations derived from the EP
and GTD experiments are simultaneously consistent with the results of all of
the local and NL experiments used for testing gravitational theories.
According to them:
\end{itemize}

\begin{enumerate}
\item  The G field is a space with a gradient of the NL speed of light.

\item  The basic NL parameters of the bodies at rest in different positions
of a static G field are well-defined functions of their NL positions in the
field. They are proportional to the square root of the NL speed of light.

\item  {\it During a free fall in a static field, the NL mass-energy of a
body, with respect to any well-defined observer in a fixed potential,
remains constant (NL mass-energy conservation during a free fall.).}

\item  {\it A static G field does not exchange energy either with radiation
or with the test bodies. }

\item  {\it It is the body, not the field, the one that puts on the energy
for the G work.}
\end{enumerate}

The G energy is a fraction of the mass-energy of the particles which is
released by the effect of the lower NL speed of light that exists in lower G
potentials. In different words, gravity is more related to matter
annihilation than to the energy given up by some external force.

{\it These conclusions are in clear discrepancy with the Einstein hypothesis
on the energy exchange between the G field and the bodies.}

The above results have also been explained from a non-conventional approach
based on a more general form of the EP. According to it, {\it all of the
well-defined parts of a local system must obey the same general physical laws%
}. {\it Thus any quantum of radiation in stationary state must have the same
inertial and gravitational properties as ordinary matter}.

Consequently, the general properties of matter can also be derived from the
theoretical properties of a particle model made up of radiation in
stationary state. Thus the acceleration of gravity comes from a refraction
phenomenon that does not change the average frequency of the photons. This
accounts, globally, for the no-exchange law derived from the experiments.

According to the nature of particle model, its G field turns out to be the
result of interference of wavelets with ``random phases''. Then {\it the
long-range G field itself has no energy because the net wave amplitude is
zero}. Thus the probability for the existence of energy in a G field is also
zero.\label{21}

Then the experimental facts{\it \ rule out the possibility of existence of
some appreciable G field energy that can be exchanged with the bodies or
particles}''. This would question the possibility for the existence of
gravitons of properties similar to those of photons. However, the most exact
experiments are not exact enough to discard the possibility for the
existence of some much weaker kind of energy exchange between photons, or
bodies, and G fields.

The above errors have prevailed for about a century because they are tacitly
supported by the classical-like hypothesis in that matter does not change
after a change of G potential with respect to an observer. Then most people
do not realize that the relations between quantities referred to standards
located in different potentials, are inhomogeneous and without well-defined
physical meaning. In the current literature such inhomogeneity errors are
not obvious because most of the times the positions of the reference
standards are not explicitly stated in each quantity.

Due to the lack of energy of the G field,  new kind of black hole and new
universe must have some radical differences with the conventional ones\cite
{4Vera81a,5Vera81b,6Vera97,12Vera83,13Vera86,15Vera99a}. . More details are
given in 
\mbox{$<$}%
http://sites.netscape.net/rafaelveram/index.htm%
\mbox{$>$}%
and 
\mbox{$<$}%
http://educar.org/cecc/rvera/fotonu\_00.htm%
\mbox{$>$}%
.

\subsection{Acknowledgments}

I wish to thank to Remo Ruffini, and Walter Thirring for friendly
encouragement during the Einstein Centennial Symposium on Fundamental
Physics in 1979. To Thomas B. Andrews, Patricio Diaz and my colleagues, for
helpful assistance and friendship.


\begin{references}
\bibitem{1Einstein}  A. Einstein, {\it The Meaning of Relativity} (Princeton
University Press, New Jersey, 1955), pp 82--83

\bibitem{2Vessot}  F. Vessot {\it et al, Phy. Rev. Lett. }{\bf 45, }pp
2081-84 (1980).

\bibitem{3Misner}  Ch. W. Misner, K. S. Thorne and J. A. Wheeler, {\it %
Gravitation} (Freeman, San Francisco, 1973), p 386

\bibitem{4Vera81a}  R. A. Vera, in\ {\it Proceedings of The Einstein
Centennial Symposium on Fundamental Physics}, Bogot\'{a}, 1979, edited by S.
M. Moore, A. M. Rodriguez-Vargas and G Violini. (Universidad de Los Andes,
Bogot\'{a}, 1981), pp. 597--625.

\bibitem{5Vera81b}  R. A. Vera, {\it Int. J. Theor. Phy,} (Plenum Publishing
Corporation, 1981)\footnote{%
(It is available in 
\mbox{$<$}%
http://sites.netscape.net/rafaelveram/index.htm%
\mbox{$>$}%
).}, {\bf 20}, pp 19-50.

\bibitem{6Vera97}  R. A. Vera, {\it The New Universe Fixed by the
Equivalence Principle and Light Properties} (Ediciones Universidad de
Concepcion. Chile. 1997)\footnote{%
(This book is available in 
\mbox{$<$}%
http://sites.netscape.net/rafaelveram/index.htm%
\mbox{$>$}%
).}

\bibitem{7Pound}  R. V. Pound and J. L. Snider, {\it Phy. Rev. B}, {\bf 140,}
pp 788-803 (1965)

\bibitem{8ShafRad}  I. I. Shapiro {\it et al, Phy. Rev. Lett,} {\bf 26}, pp
1132-1135 (1971)

\bibitem{9BertotiRef}  B. Bertoti, D. Brill and R. Krotkov, in\ {\it %
Gravitation: An Introduction to Current Research}, ed. L. Witten, (Wiley,
NY, 1962)

\bibitem{10ShafPeri}  I. I. Shapiro {\it et al}, {\it Phys. Rev. Let}, Vol 
{\bf 28}, pp.1594-1597 (1972)

\bibitem{11Kennedy}  R. J, Kennedy and Thorndike, {\it Phy. Rev}. {\bf 42},
400 (1932)

\bibitem{12Vera83}  R. A. Vera, Introducci\'{o}n a una teor\'{i}a nolocal de
la f\'{i}sica aplicada a campos gravitacionales, Departamento de F\`{i}sica.
Universidad de Concepci\'{o}n Chile, Report No 1, 1983

\bibitem{13Vera86}  R. A. Vera, in {\it Proceedings of the Fourth Marcel
Grossmann Meeting on General Relativity}, Roma, 1985, editted by R. Ruffini
(Elsevier Science Publishers V. B., 1986), pp. 1743-1752

\bibitem{14Vera96a}  R. A.Vera, in {\it Proc. of the 7th Marcel Grossmann
Meeting on General Relativity}, Stanford, 1994, edited by. R. T. Jantzen and
G. Mac Keiser and R. Ruffini. (World Scientific, Singapore, 1996), Vol A,
pp. 511--513.

\bibitem{15Vera99a}  R. A. Vera, in {\it Proc. of the 8th Marcel Grossmann
Meeting on General Relativity}, Jerusalem, 1997, edited by Tsvi Piran and R.
Ruffini, (World Scientific Publishing Co. Singapore, 1999), Vol A, pp.
303-306
\end{references}
\end{document}